\documentclass[%
reprint,
twocolumn,
superscriptaddress,
graphicx,
floatfix,
amsmath,
amssymb,
bibnotes,
aps,
prl,
]{revtex4-2}
\usepackage[]{graphicx}
\usepackage{epstopdf}
\usepackage{MnSymbol}
\usepackage{latexsym}
\usepackage{color}
\usepackage{dcolumn}
\usepackage{bm}
\usepackage[mathlines]{lineno}
\usepackage{usebib}

\begin{document}


\title{Plasmon-enhanced ultrafast time-resolved spectroscopy of NV-containing diamond}

\author{Yuta Kimura}
\affiliation{Department of Applied Physics, University of Tsukuba, 1-1-1 Tennodai, Tsukuba 305-8573, Japan.}

\author{Toshu An}
\affiliation{School of Materials Science, Japan Advanced Institute of Science and Technology, Nomi, Ishikawa 923-1292, Japan.}

\author{Muneaki Hase}
 \email{mhase@bk.tsukuba.ac.jp}
\affiliation{Department of Applied Physics, University of Tsukuba, 1-1-1 Tennodai, Tsukuba 305-8573, Japan.}

\begin{abstract}
We investigated ultrafast nonlinear optical effects in nitrogen-vacancy (NV)-containing diamond which is in contact with a gold-coated blazed diffraction grating using a pump-probe reflectivity technique.
The reflectivity change caused by optical Kerr effect and two-photon absorption was enhanced several times because of the electric field enhancement induced by the propagating surface plasmon (PSP).
Furthermore, by performing measurements with varying the incident angle of the pump beam and numerical simulations of the electric field using the Finite Difference Time Domain method, signal enhancement due to the PSP was confirmed both experimentally and theoretically. This study paves the way for applications based on enhanced nonlinear optical effects in diamond.
\end{abstract}

\date{\today}

\maketitle
The nitrogen-vacancy (NV) center in diamond is a promising system for a wide range of quantum technologies, such as quantum computing\cite{pezzagna2021quantum}, quantum information processing\cite{prawer2014quantum}, and quantum sensing\cite{degen2017quantum}.
Particularly in the field of quantum sensing, it allows for highly sensitive measurements of the magnetic field\cite{taylor2008high}, electric field\cite{dolde2011electric}, and even the temperature\cite{neumann2013high}. Significant progress has been made in their practical applications, including biomedical applications\cite{aslam2023quantum} and semiconductor device testing\cite{iwasaki2017direct}.
However, conventional sensing technologies are limited by a time resolution of around a microsecond to a nanosecond\cite{dolde2011electric}, preventing the measurement of changes in electric and magnetic fields occurring on timescales less than a picosecond. 
As a result, recent research has begun to focus on ultrafast nonlinear optical responses of the NV centers in diamonds\cite{motojima2019giant,KUDRYASHOV2025112081}.
Diamonds possess inversion symmetry, resulting in second-order nonlinear susceptibility $\chi^{(2)}= 0$ (Ref.\cite{Shen2003principles,Trojanek2010NLO}). However, the introduction of NV centers into diamonds breaks the inversion symmetry, leading to $\chi^{(2)}\ne 0$ (Ref.\cite{abulikemu2021second,Ichikawa2024CP}).
This property has allowed the observation of second-order nonlinear phenomena, such as second harmonic generation (SHG)\cite{abulikemu2021second,abulikemu2022temperature}, the cascading optical Kerr effect\cite{motojima2019giant}, and the inverse Cotton-Mouton effect\cite{sakurai2022ultrafast}, which cannot be observed in pure diamonds\cite{Almeida2017OKE}.
Studying the nonlinear characteristics that change by introducing NV centers into diamonds could pave the way for new applications, including quantum frequency conversion\cite{han2021microwave}, terahertz sources\cite{lewis2014review}, ultrafast optical switching\cite{chai2017ultrafast}, and ultrafast field sensors\cite{zhang1996ultrafast}, expanding beyond the traditional uses of NV centers.

However, to detect ultrafast nonlinear optical responses, it is necessary to irradiate the diamond with high-intensity laser pulses 
($\geq$100 GW/cm$^2$)\cite{motojima2019giant,sakurai2022ultrafast}. Consequently, for practical applications, it is crucial to enhance signal strength to enable adequate signal detection with compact and lower-intensity lasers, such as a semiconductor laser\cite{agrawal2013semiconductor}. Therefore, we have focused on the surface plasmon, which enhances light matter interaction, including fluorescence\cite{li2017plasmon}, Raman scattering\cite{wang2020fundamental}, infrared absorption\cite{osawa1993surface}, and nonlinear optical effects\cite{kauranen2012nonlinear} through its electric field enhancement.
The surface plasmon refers to the collective oscillation of electrons at the interface between a metal and a dielectric \cite{Mulvaney1996SP} and it has been extensively investigated since 1902 when Wood anomaly was observed with metallic diffraction gratings \cite{Wood1902}. It can be broadly classified into two types: the localized surface plasmon (LSP)\cite{Mayer2011LSP} and the propagating surface plasmon (PSP)\cite{Barnes2003SPP}.
Although LSP achieves an electric field enhancement larger than that of PSP, its effective use demands more precisely fabricated structures.

In this study, we investigate the enhancement of the ultrafast nonlinear optical response in NV-containing diamond using PSP, which may provide a simple approach to signal enhancement.
Among nonlinear optical effects, we evaluated the optical Kerr effect (OKE) and two-photon absorption (TPA), involving a cascade process induced by $\chi^{(2)}$ interactions from the NV centers, using time-resolved pump-probe spectroscopy.
A simple configuration where the NV-containing diamond is in contact with a gold-coated blazed diffraction grating was adopted to apply the PSP to the amplification of the ultrafast nonlinear optical response.
A gold-coated blazed diffraction grating was utilized as the metal structure for exciting the PSP.
The optimal incident angle of light for the excitation of the PSP at the measurement frequency was simulated using the Finite Difference Time Domain (FDTD) method\cite{Allen1995FDTD} for the metallic structure.
As a result, by adjusting the incident angle of the pump light that induces nonlinear optical effects, we were able to enhance the ultrafast nonlinear optical response several times.
Furthermore, by varying the incident angle of the pump light during the measurements, we were able to obtain an incident angle dependence that was consistent with the FDTD simulation.

The diamond sample was the same as that used in the previous study\cite{motojima2019giant}; a [100] type-IIa single crystal made by Element Six using the CVD method, and its size was 3.0 mm × 3.0 mm × 0.3 mm.
30 keV nitrogen ions ($^{14}$N$^+$) were implanted into one surface of the diamond at the N$^+$ dose levels of $1.0 \times 10^{12}$ ions/cm$^2$ to introduce NV centers in the diamond. 
The implant depth deduced from Monte Carlo simulations was approximately 30-40 nm\cite{kikuchi2017long} and the profile was close to a Gaussian function with FWHM of $\sim$ 50 nm.
In other words, it can be said that the NV centers exist below one side of the diamond's surface.
The diamond was then annealed at 900-1000 $^\circ$C in an argon atmosphere for an hour to create NV centers with a production efficiency of $\sim 1 \%$ (Ref.\cite{pezzagna2010creation}).
The estimated density of the NV centers was $\sim 2.0 \times 10^{17}$/cm$^3$.

\begin{figure}[h]
   \begin{center}
  \includegraphics[width=8.8cm]{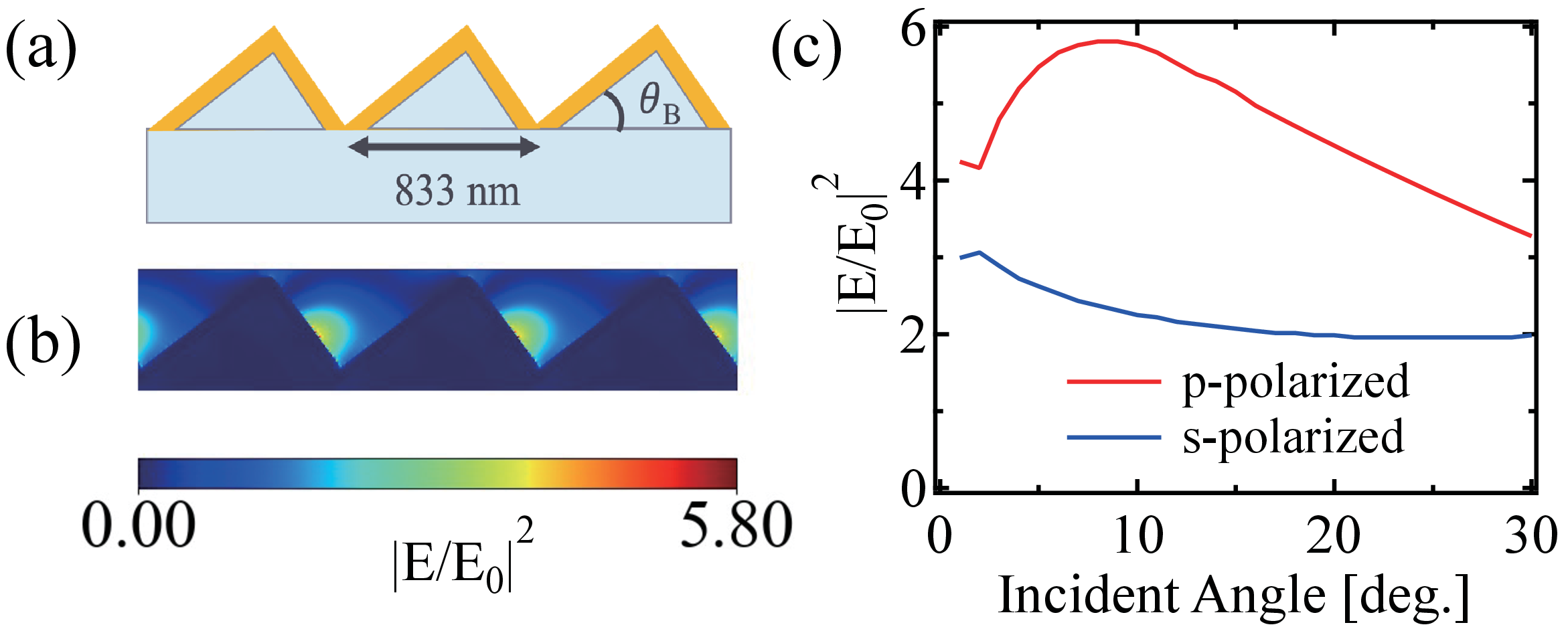}
  \caption{(a) The schematic structure of the gold-coated brazed diffraction grating with a blaze angle of $\theta_{B}=36.87^{\circ}$. (b) Electric field distribution $|E/E_0|^2$ for the p-polarized Gaussian source at incident angle of $\theta=9^{\circ}$ simulated by the FDTD method. (c) The FDTD simulation results of maximum electric field enhancement $|E/E_0|^2$ as a function of incident angle $\theta$.}
  \label{fig.1}
     \end{center}
\end{figure}

In addition, to enhance the nonlinear optical effects of NV-containing diamond by surface plasmon, we used a gold-coated blazed diffraction grating made by Edmund Optics. Gold is coated on the floating glass substrate diffraction grating with a blaze angle of $\theta_{B}=36.87^{\circ}$ and a groove density of 1200 lines/mm (see Figure \ref{fig.1}).

The enhancement effect of the gold-coated blazed diffraction grating was confirmed through FDTD simulations with MIT Electromagnetic Equation Propagation (MEEP)\cite{oskooi2010meep}.
The blazed angle was set at $\theta=36.87^{\circ}$, the groove density was 1200 lines/mm, and the thickness of the gold coating was 50 nm.
The computational domain was discretized into 5 nm × 5 nm cells, applying periodic boundary conditions in the direction of the periodic structure and PML (perfectly matched layer) boundary conditions in the perpendicular direction.
Gaussian sources with p- and s-polarization, both with a central wavelength of 800 nm (determined to meet the conditions of subsequent experiments), were incident on the diffraction grating. The simulation was repeated by varying the incident angle by 1 degree at a time to calculate the angle dependence of the maximum electric field enhancement due to the gold-coated blazed diffraction grating.

\begin{figure}[h]
   \begin{center}
  \includegraphics[width=8.8cm]{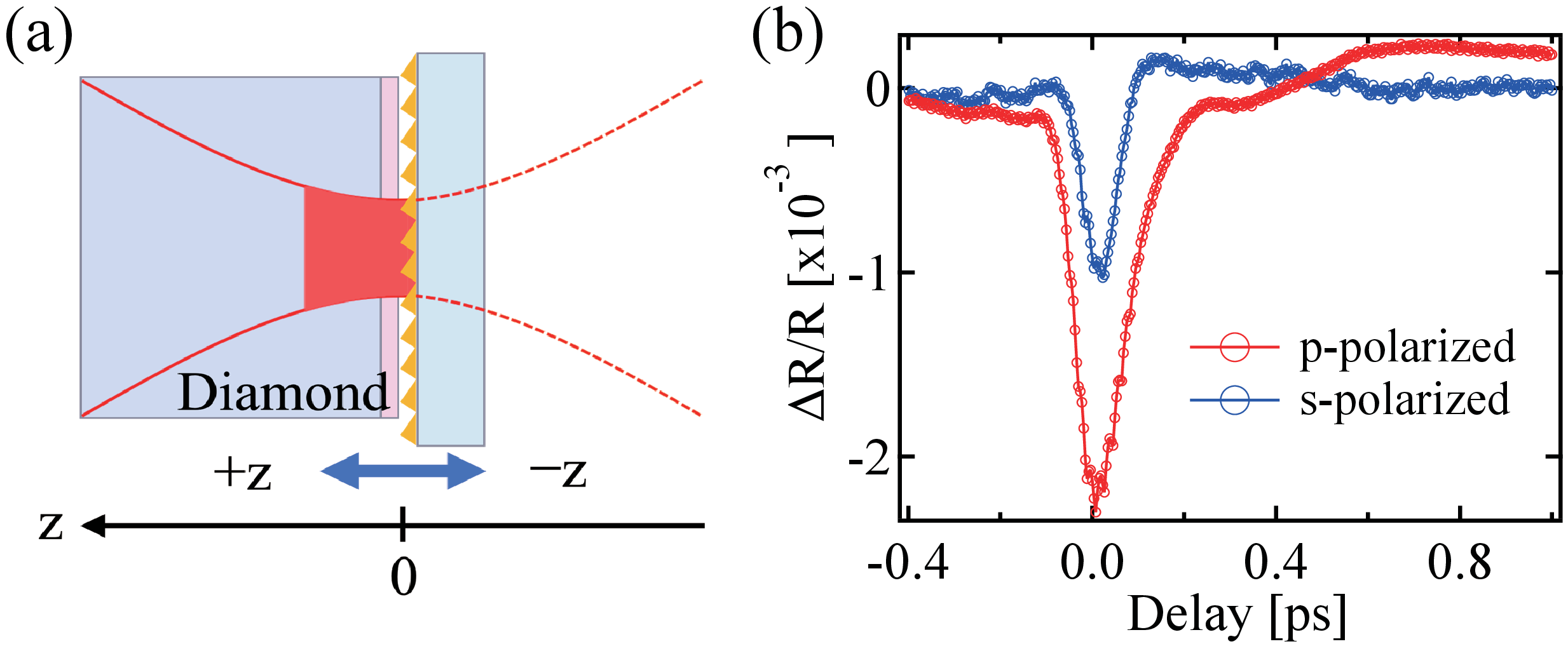}
  \caption{(a) Schematic illustration of the experimental configuration where the NV-containing side of the diamond is in contact with the diffraction grating, and the laser focal spot is aligned with the contact region. (b) The results of pump-probe reflectivity measurement at $z$ = 0 $\mu$m where the signal is maximized under p-polarized pump excitation.}
  \label{fig.2}
     \end{center}
\end{figure}

Furthermore, to enhance the nonlinear optical effects in the NV-containing diamond through the PSP, we took advantage of the fact that the NV centers are located on one surface of the diamond. NV-containing side was placed in contact with the diffraction grating, and laser pulses were irradiated from the opposite side.
This simple measurement setup allows us to explore the excitation of PSP and the associated enhancement of nonlinear optical responses from the NV-containing diamond by aligning the laser focal spot with the region where the NV centers and the diffraction grating are in contact, as shown in Fig.\ref{fig.2}(a). When the diamond and the diffraction grating are shifted backward, the focal spot moves into a region lacking NV centers, and we observe signals from the pure diamond region without PSP enhancement. Conversely, moving the sample forward causes the laser pulses to reflect off the grating surface, and the focal spot again overlaps with a pure diamond region, from which a similarly unenhanced signal is obtained.

To demonstrate this concept, we performed pump-probe reflectivity measurements while varying the position of the diamond and the diffraction grating relative to the laser focal spot. The laser pulses whose pulse width was $\sim$ 40 fs, central wavelength 800 nm, repetition rate 100 kHz, and average power $\geq$ 500 mW were emitted from a regenerative amplifier system (RegA9040). The pump and probe pulses were co-focused with a $f = 150$ mm lens, and the calculated spot size of the pump pulse was 73 $\mu$m. To obtain a time-resolved reflectivity signal, the optical path length of the pump beam was scanned at 10.0 Hz using a shaker. In addition, we measured the reflectivity change as a function of the incident angle of the p-polarized pump pulses $\theta$ from $2^{\circ}$ to $17^{\circ}$ to compare with the simulation results. In both experiments, deviation in the propagation direction of the probe beam caused by changes in the sample position or angle was compensated by adjusting the position of the detector using a micrometer stage. 

Figure \ref{fig.1}(b) shows the simulation result of the distribution of the electric field $|E/E_0|^2$ when p-polarized light was irradiated on a gold-coated blazed diffraction grating at an incident angle of $\theta = 9^{\circ}$.
The large electric field enhancement near the gold surface is due to the excitation of the PSP, which causes the field-enhancement effect.

Although another previous study\cite{wang2017plasmon} has shown that the enhancement of the electric field is highest at the apex of the diffraction grating, Fig.\ref{fig.1}(b) reveals that the enhancement is highest in the groove region. This difference is attributed to the direction from which the light is incident, either from the long side or the short side of the blazed diffraction grating.
According to a previous study that enhanced nonlinear optical effects using surface plasmon generated by a diffraction grating\cite{genevet2010large}, the electric field enhancement at the grating surface influences the emission intensity and directionality of the light emitted from the grating side (referred to as out-coupling in Ref.\cite{genevet2010large}).
Therefore, in the system that will be discussed later, where the diffraction grating is in contact with a diamond, the light reflected by the diffraction grating is considered to be influenced by the electric field enhancement due to the surface plasmon, which in turn enhances the nonlinear optical response in the diamond as the light passes through the diamond again.

Figure \ref{fig.1}(c) presents the dependence of the maximum electric field enhancement factor $|E/E_0|^2$ on the incident angle for both p- and s-polarized light. No peak is observed in the case for s-polarized light, as it cannot couple with the PSP. 
In contrast, the result for the p-polarized light exhibits a peak at $\theta = 9^{\circ}$, where the electric field enhancement $|E/E_0|^2$ is $\approx$2.5 times greater than for the case of the s-polarized light.
Therefore, the optimal incident angle for p-polarized light with a wavelength of 800 nm is $\theta = 9^{\circ}$.

Next, we performed pump-probe reflectivity measurements for the system in which the diamond was placed in contact with the diffraction grating and translated back and forth as shown in Fig.\ref{fig.2}(a). 
Given that the diamond is 300 $\mu$m thick, we varied the position of the sample from the origin position ($z$ = 0 $\mu$m) in steps of 100 $\mu$m within a range of $\pm$500 $\mu$m to examine the variation of the signal in detail. Here, the origin position ($z$ = 0 $\mu$m) was determined by the surface of the grating (or the back surface of the diamond sample). 
At the surface of the diffraction grating, the influence of electric field enhancement is significant, and in the previous study\cite{motojima2019giant}, it has been shown that NV-containing diamond exhibits signal intensity stronger than pure diamond because NV centers induce a breaking of the inversion symmetry in the diamond lattice, allowing $\chi^{(2)}\ne 0$ and improving nonlinear optical responses through cascade processes.
Therefore, the signal is expected to be maximized when the laser focal point overlaps with the NV center-grating interface, as shown in Fig.\ref{fig.2}(a). The pump pulses were incident at an angle of approximately $9^{\circ}$, and measurements were performed with both p- and s-polarized pump pulses. The probe polarization was set to be orthogonal to that of the pump.

 Figure \ref{fig.2}(b) shows the result of the pump-probe reflectivity measurement at $z$ = 0 $\mu$m, where the signal was maximized under p-polarized pump excitation. The results for p-polarization and s-polarization showed a sharp decrease. 
Their FWHM was estimated to be 128.3 $\pm$ 2.4 fs for the p-polarization and 86.4 $\pm$ 1.8 fs for the s-polarization, by Gaussian fitting. The difference in FWHM is likely due to variations in the contribution of nonlinear optical effects induced by the electric field enhancement. The reflectivity change $|\Delta R/R|$ is considered to be induced by nonlinear optical effects, such as OKE and TPA \cite{motojima2019giant}. 
The refractive index change due to OKE and the absorption coefficient change due to TPA are described by
$n = n_0 + n_2I$, $\alpha = \alpha_0 + \beta I$, respectively, where $I$ is the light intensity, $n_0$ and $n_2$ represent the linear and nonlinear refractive index, respectively, and $\alpha_0,\beta$ represent the linear and nonlinear absorption coefficient, respectively.
Applying this absorption coefficient to the Beer-Lambert law shows that the number of free carriers generated by TPA scales with the square of $I$\cite{rumi2010two}. Furthermore, based on the Drude model\cite{downer1986ultrafast}, the refractive index change induced by the generated free carriers is also expected to scale quadratically with $I$, in contrast to the refractive index change caused by OKE, which is linear with respect to $I$\cite{motojima2019giant}.
As a result, it is considered that the electric field enhancement effect further amplified the contribution from the TPA. 
Since the relaxation of free carriers typically takes longer ($\sim$ several hundred fs) than the response time of OKE, the signal under p-polarized excitation is expected to exhibit a broader temporal profile.

It should be noted that the slow dynamics of the transient reflectivity (0.2-1.0 ps) will possibly be induced by thermal effects and is larger for the case of p-polarization than that of s-polarization. The temperature rise would be larger for the case of the p-polarization because of the carrier-phonon thermalization after real charge carrier excitation by TPA. 

Our measurements were performed with a relatively low fluence of 3.92 mJ/cm$^2$, at which no signal was obtained in the previous study\cite{motojima2019giant}. Therefore, we applied the fitting function derived from the fluence dependence\cite{motojima2019giant} to calculate the signal intensity at the fluence of 3.92 mJ/cm$^2$, yielding $|\Delta R/R|$ = 1.35$\times 10^{-4}$.
If we compare this $|\Delta R/R|$ value with the highest enhanced $|\Delta R/R|$ signal  in our measurement $|\Delta R/R|$  = 2.30$\times 10^{-3}$ as shown in Fig.\ref{fig.2}(b), the signal enhancement can be estimated $\sim$ 17 times. 
However, it should be noted that, unlike the previous study\cite{motojima2019giant} in which the laser beam was irradiated from the NV-center side of the diamond, the present study involves irradiation from the opposite side. Thus, further experiments are required to accurately evaluate the enhancement factor, such as fabricating a grating structure directly on the NV-center side of the diamond. Even though, a comparison between the results for the p- and s-polarized light in Fig.\ref{fig.2}(b) reveals that the signal is enhanced by a factor of $\approx$2.2. This enhancement is in good agreement with the simulation comparing the results for p- and s-polarized light at an incident angle of $\theta=9^{\circ}$ (see Fig. 1). This indicates that the PSP-induced enhancement predicted by the simulation was indeed realized in our simple setup, where the diamond is in contact with a gold-coated blazed diffraction grating.

\begin{figure}[h]
   \begin{center}
  \includegraphics[width=7.0cm]{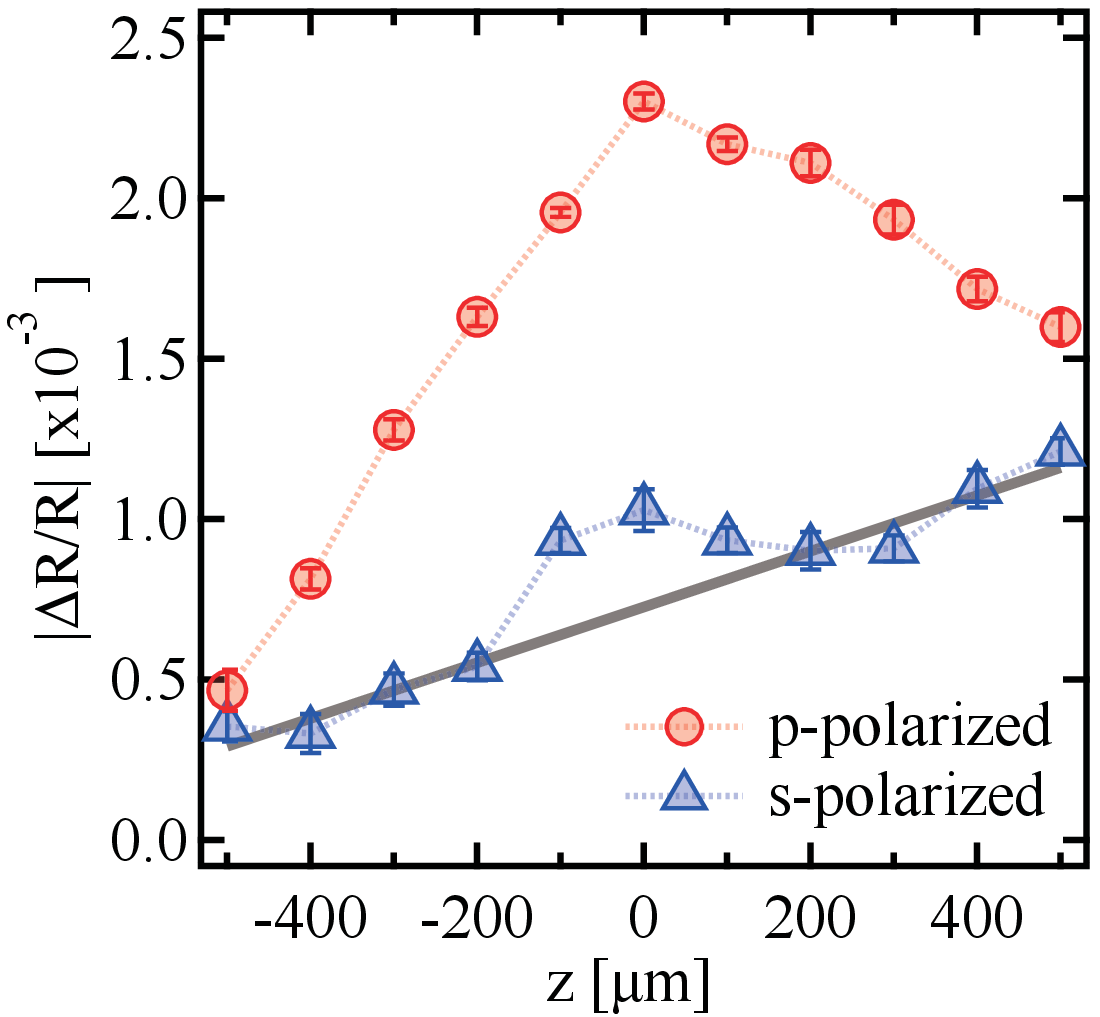}
  \caption{The sample position ($z$) dependence of the pump-probe reflectivity measurements, relative to the position of maximum signal under p-polarized pump excitation. The gray solid line represents a linear fit to the result of the s-polarization, excluding the values at $z$ = 0, $\pm$100 $\mu$m.}
  \label{fig.3}
     \end{center}
\end{figure}

Figure \ref{fig.3} shows the sample position dependence of the absolute value of the reflectivity change $|\Delta R/R|$. 
A gradual decrease in signal intensity is observed under p-polarized pump excitation as the sample is translated away from $z$ = 0 $\mu$m.
This is attributed to the reduction in PSP-induced signal enhancement due to the displacement of the sample.
Note that while PSP is confined to a region near the grating surface on the order of the optical wavelength, the signal from p-polarization remains higher than that from s-polarization even when the sample is shifted by about $\pm$400 $\mu$m. In our setup, the Rayleigh length $z_R$ was calculated to be 5.3 mm using the equation $z_R = \pi w_0^2 / \lambda$, where $w_0$ is the beam waist, $\lambda$ is the wavelength, respectively. The pump and probe beams were co-focused and the angle between them was estimated to be $\sim 5^{\circ}$, derived from a lens with a 0.5-inch radius and a focal length of 150 mm. Assuming that the laser beams overlap as cylinders with a diameter of $\approx$73 $\mu$m, their spatial overlap length is $\approx$840 $\mu$m. This suggests that the effective detection region for nonlinear optical signals is long enough to capture the signal from the vicinity of the diffraction grating even when the focal spot is located several hundred micrometers away from the NV-grating interface, thanks to PSP-induced enhancement.

In the case of s-polarized excitation, an increase in signal intensity at $z$ $\approx$ 0 is observed on a linear background (see the gray solid line in Fig.\ref{fig.3}). The increase at $z$ $\approx$ 0 can be attributed to the detection of the NV-containing region, resulting in a higher signal intensity \cite{motojima2019giant}.
\color{black}
When the sample is shifted by several hundred micrometers without PSP enhancement, the focal spot no longer overlaps with the NV-containing region where the signal is stronger, and only the pure diamond region where the signal is weaker is probed.

By comparing the signal intensity at $z$ = 0 with the background value estimated from the linear fit, the enhancement factor due to the NV-containing region is found to be $\sim$ 1.4. This result is in good agreement with previous z-scan measurements\cite{motojima2019giant}, which reported enhancement factors of 1.3 and 1.6 for the nonlinear refractive index and the two-photon absorption coefficient, respectively, when comparing the pure diamond with the NV-containing diamond.
These results suggest that the NV-containing region was detected at $z$ $\approx$ 0.
The linear background is also observed in the p-polarized case, leading to an asymmetry with respect to $z$ = 0. 
It can be attributed to the interference effect\cite{jiang2011surface} of the diffraction grating, which enhances the reflected pump pulses.
As $z$ increases, the contribution of pump pulses reflected by the diffraction grating becomes more significant, leading to an increased influence of enhancement due to the interference effect. This likely contributes to the linear background.

\begin{figure}[h]
   \begin{center}
  \includegraphics[width=8.0cm]{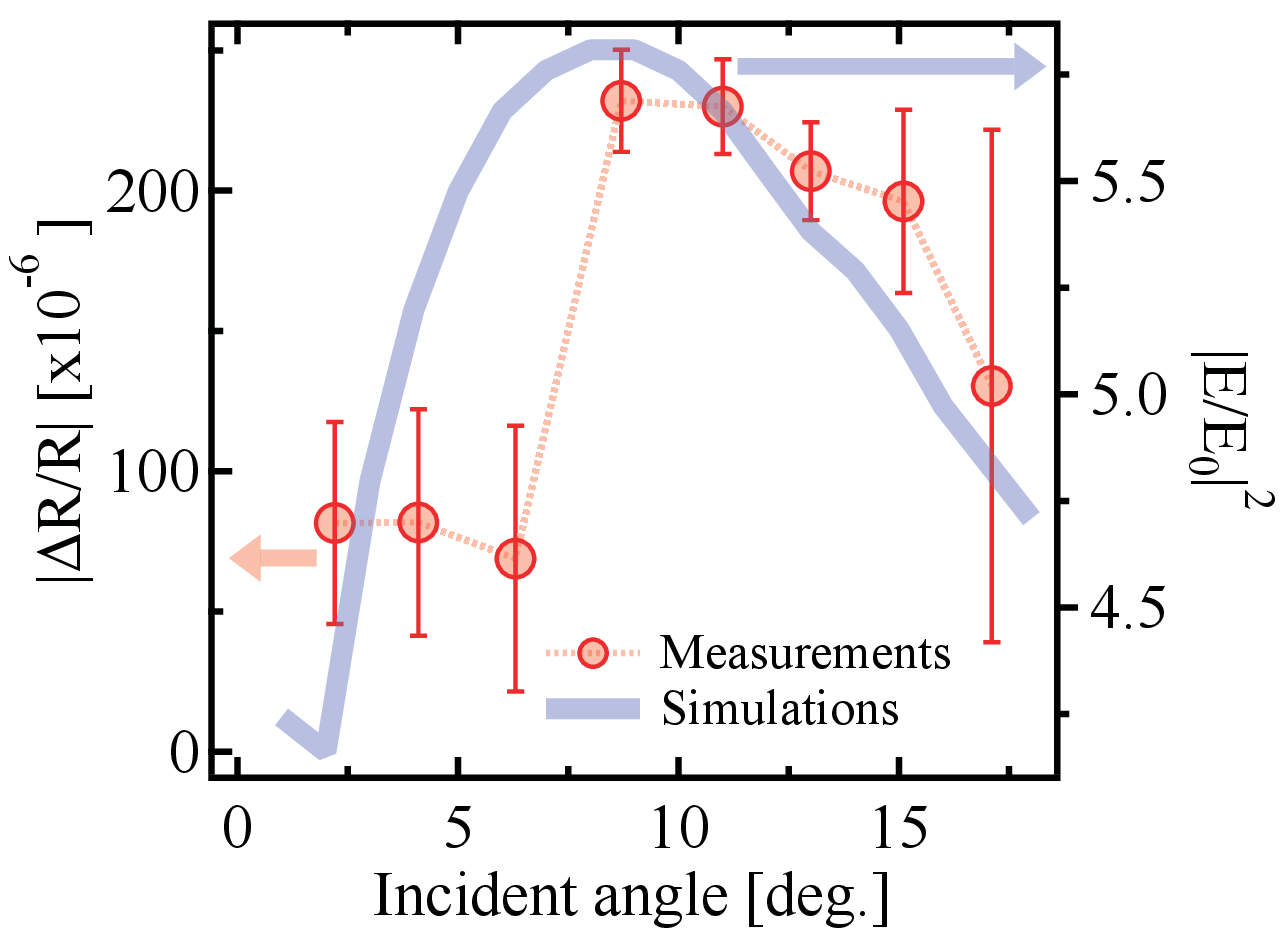}
  \caption{The incident angle ($\theta$) dependence of reflectivity change $|\Delta R/R|$ induced by the p-polarized pump pulses together with the FDTD simulation results of maximum electric field enhancement $|E/E_0|^2$.}
  \label{fig.4}
     \end{center}
\end{figure}

Moreover, we measured the reflectivity change as a function of the incident angle of the p-polarized pump pulses $\theta$ from $2^{\circ}$ to $17^{\circ}$ at z $\approx$ 0 to confirm that the signal enhancement was caused by PSP in more detail.
Figure \ref{fig.4} shows the results of such measurements together with the results of the FDTD simulations.
There is an enhanced peak at $\theta = 9^{\circ}$ corresponding to the simulation of electric field enhancement $|E/E_0|^2$.
Due to the long spatial overlap of the pump and probe beams, the enhancement effect of PSP was observed over a broad range in position-dependent measurements. However, in angle-dependent measurements, good agreement between the experimental results and the simulation confirmed that the PSP plays a crucial role in signal enhancement in our experimental setup.

The signal intensity of the reflectivity change $|\Delta R/R|$ varied $\approx$2.8 times with change in angle, while the enhancement of the electric field in the simulation changed only by a factor of 1.4 when we take the value of its minimum at $\theta = 2^{\circ}$ and its maximum at $\theta = 9^{\circ}$, being $\approx$1/2 for the case of the experiment. 
One possible reason for this discrepancy is that the confinement of light between the diffraction grating and the diamond was significantly increased, leading to a stronger electric field enhancement effect than in the simulation. 

In addition, the refractive index change caused by TPA may also play a role because it is proportional to the square of the light intensity.
The previous study\cite{motojima2019giant} has suggested that OKE predominantly influences the signal. However, in the present measurement, the refractive index change caused by TPA, which is more strongly affected by the electric field enhancement, may no longer be negligible.

Since the measurement was performed by varying the sample angle, the angle of the probe beam also changed accordingly. Due to the difference in the incidence angles between the pump and probe beams $\sim 5^{\circ}$, the refraction angles inside the diamond differ slightly. However, the resulting shift of spatial overlap between the pump and probe beams when the sample angle is varied from $2^{\circ}$ to $17^{\circ}$ relative to the pump beam is estimated to be $\sim $ 300 nm, which is negligible compared to the beam spot size of 73 $\mu$m.

Because the simulation of the electric field enhancement was performed only for the gold-coated blazed diffraction grating without the diamond, it does not fully reproduce the experimental results obtained for the system in contact with the NV-containing diamond.
Therefore, more precise simulations are required to achieve better agreement between the experimental and simulated results.

In conclusion, this study explored the ultrafast nonlinear optical effects of NV-containing diamond through a pump-probe reflectivity technique. By placing the NV-containing side of the diamond on a gold-coated blazed diffraction grating and focusing a laser pulse on it, we obtained the signal enhanced several times due to PSP.
Position-dependent measurements revealed the contribution from the NV centers and the signal modulation due to the PSP.
Angle-dependent measurements were consistent with FDTD simulations, confirming that PSP-induced electric field enhancement is key to boosting the nonlinear optical effects in NV-containing diamond.
This study paves the way for applications based on enhanced nonlinear optical effects in diamond.


\begin{acknowledgments}
We acknowledge funding support from Japan Society for the Promotion of Science (23K22422, 24K01286, 25H00849); Core Research for Evolutional Science and Technology program of the Japan Science and Technology (JPMJCR1875). 
The authors acknowledge Siya Deshpande for critical reading of the manuscript and useful feedback. 
\end{acknowledgments}

\bibliography{a_ref}

\end{document}